\documentclass[aps,preprint,onecolumn,nofootinbib]{revtex4-1}

\usepackage{amsmath}
\usepackage{amssymb}
\usepackage{graphicx}
\usepackage{hyperref}
\graphicspath{{./figures/}}

\newcommand{\degC}{^{\circ}\mathrm{C}}
\newcommand{\cmc}{\ensuremath{\mathrm{cmc}}}

\begin{document}
\title{Temperature dependence of surface tension in aqueous sodium dodecyl sulfate solutions}
\author{Jong Cheon Lee}
\affiliation{Department of Mechatronics, Konkuk University, Chungju 27478, South Korea}
\author{Ildoo Kim}
\email{ildookim@kku.ac.kr}
\affiliation{Department of Mechatronics, Konkuk University, Chungju 27478, South Korea}

\begin{abstract}
We report measurements of the surface tension of aqueous sodium dodecyl sulfate solutions at sixteen concentrations from 0.17 to 43\,mM, spanning 0.02 to 5 times the room-temperature critical micelle concentration (\cmc), and at temperatures from 26 to 83\,$\degC$, using the Wilhelmy plate method.
At the lowest concentration the surface tension decreases with temperature, as it does for pure water.
Above about 0.4\,mM, however, it increases: the temperature coefficient changes sign at a concentration of 0.3 to 0.8\,mM, roughly a tenth of the \cmc, and reaches $+0.53\,\rm mN\,m^{-1}\,K^{-1}$ near 3\,mM, three times the magnitude of that of pure water and of the opposite sign.
Above the \cmc\ the surface tension is nearly independent of temperature, saturating at $35.2\pm0.3\,\rm mN\,m^{-1}$.
The behavior cannot be attributed to the temperature dependence of the \cmc, which is an order of magnitude too weak.
It is instead consistent with a surfactant layer that thins as the temperature rises.
The results imply that the sign and magnitude of thermocapillary stresses in SDS solutions can be selected through the concentration.
\end{abstract}
\maketitle

%%%%%%%%%%%%%%%%%%%%%%%%%%%
\section{Introduction}
%%%%%%%%%%%%%%%%%%%%%%%%%%%

%The complexity of fluid dynamics comes not only from its nonlinear and singular nature but also from implicit and inconspicuous correlation between variables.

The surface tension $\sigma$ of surfactant-laden interfaces is usually characterized by the surface equation of state
\begin{equation}
\sigma= \sigma_0 - RTf(\Gamma),
\label{eq:surface_tension_linear}
\end{equation}
where $\sigma_0$ is the surface tension of the pure liquid, $R$ the ideal gas constant, $T$ the temperature, $\Gamma$ the surface concentration of surfactant, and $f$ a function that is specific to the system.
Eq. \eqref{eq:surface_tension_linear} is obtained by integrating the two-dimensional Gibbs--Duhem equation
\begin{equation}
S\,dT+A\,d\sigma+ N_i\,d\mu_i = 0
\label{eq:Gibbs-Duhem}
\end{equation}
at fixed temperature, i.e., $dT=0$.
Therefore, Eq. \eqref{eq:surface_tension_linear} is the surface tension {\it isotherm}: $\sigma_0$, $f$, and $\Gamma$ all carry their own implicit temperature dependence.
If this were overlooked, one might naively conclude that $\sigma$ becomes zero at a high enough temperature.

In this regard, the temperature dependence of the surface tension of surfactant-laden fluids remains in the empirical realm.
Characterizing this dependence is crucial for thermocapillary flows such as those in soap films and in wetting phenomena, but no universally applicable model exists, which makes an analytical approach difficult.
Instead, the temperature dependence of the surface tension has been investigated through experimental measurements.
For sodium dodecyl sulfate (SDS), arguably the most important surfactant in fluid dynamics, the surface tension has been measured as a function of temperature by Hern\'ainz-Berm\'udez de Castro \textit{et al.}~\cite{granada1998} (20 to 50\,$\degC$, 0.0073 to 0.14\,mM) and by Phongikaroon \textit{et al.}~\cite{judd2005} (15 to 35\,$\degC$, 0.022 to 0.117\,mM).
Both report measurements for dilute solutions, less than 1.7\% of the \cmc, and a temperature coefficient $d\sigma/dT$ indistinguishable from that of pure water.

In this study, we report the measurement of the surface tension of aqueous SDS solutions at sixteen concentrations from 0.17 to 43\,mM and at temperatures from 26 to 83\,$\degC$.
The previous reports cover only dilute solutions; measurements over the concentration range that fluid experiments actually use are missing.
We find that below the \cmc\ the surface tension increases with temperature.
The temperature coefficient reaches $+0.53\,\rm mN\,m^{-1}\,K^{-1}$ near 3\,mM, almost thrice the coefficient of pure water in magnitude but of the opposite sign.
The sign reversal occurs at a concentration of only 0.3 to 0.8\,mM, about a tenth of the \cmc, and this concentration increases with temperature.
%Above the cmc the surface tension is essentially independent of temperature, saturating at $35.2\pm0.3\,\rm mN\,m^{-1}$ over the whole range.
%We show that the temperature dependence of the cmc cannot account for these observations: collapsing our data on that assumption would require the cmc to grow fivefold between 30 and 70\,$\degC$, whereas conductometry and calorimetry give 26 to 30\,\%.
Our measurements show that increasing the temperature weakens the effect of the surfactant.

Practically, the thermocapillary response of an SDS solution can be set by its concentration.
Below the \cmc\ a thermal gradient drives Marangoni stresses opposite in direction to those in pure water; above the \cmc\ it drives almost none.
The former is a resource for thermocapillary experiments in soap films, the latter a convenience for work in which a temperature gradient is imposed for another purpose and Marangoni stresses are an unwanted complication, such as two-dimensional Rayleigh--B\'enard convection.

%%%%%%%%%%%%%%%%%%%%%%%%%%%
\section{Materials and Methods}
%%%%%%%%%%%%%%%%%%%%%%%%%%%

%%%%%%%%%%
\subsection{Apparatus}

Our measurements are carried out in a commercial tensiometer (SEO DST-30).
The tensiometer uses Wilhelmy plate method, consisting of a fixed Pt plate (width 23.85 mm, thickness 0.15 mm, total perimeter 48.00 mm) and a vertically translated stage.
On the stage, to be placed is a glass container (diameter 70 mm, height 50 mm) filled with aqueous solution of sodium dodecyl sulfate.
The stage is moved manually, and the maximum force exerted on the plate is recorded automatically.
The Pt-plate and the sample container are enclosed in a plastic chamber.

To measure the temperature of solutions, we use a grounded-junction type T thermocouple probe (Omega TJ36-CPSS-020G-4; 304 stainless sheath, 0.51 mm outer diameter, 102 mm length). 
The probe is immersed into the fluid in part, approximately 50 mm.

%%%%%%%%%%
\subsection{Solutions}

We use sodium dodecyl sulfate powder obtained from Junsei Chemical (purity 97\%) as received, without further purification.
%We use sodium dodecyl sulfate (SDS) power with purity of 97\% (TCI, cat. D1403) as received, without further purification. 
%To make a SDS solutions at various concentrations, we first make a master solution and then diluted it.
To prepare SDS solutions at various concentrations, we first prepared master solutions and subsequently diluted them with DI water.
We make 41 mM master solution by adding 11.8 g of SDS to 1000 ml of deionized (DI) water for solutions over 16.4 mM and 16.4 mM master solution for solutions under 16.4 mM.
In either case, the master solution in a sealed container is stirred overnight at room temperature.

The critical micelle concentration (\cmc) of SDS in water is a function of temperature and is 8.2\,mM at 25\,$\degC$ ($\cmc_{25}$) \cite{goddard1957,tajima1970}.
We designed the experiments to be relevant to thermocapillary flows by choosing the concentration $c_0$ of tested solutions to correspond to 0.02 to 5 times $\cmc_{25}$.
The full list of test solutions is tabulated in Table~\ref{tab:exp_conditions}.

%\begin{table}
%\centering
%\caption{Solutions examined in this work.  
%Each solution was made up at the nominal concentration $c_{0}$, heated, and then measured while cooling, so that the run begins at $T_\mathrm{max}$ with concentration $c_{i}$ and ends at $T_\mathrm{min}$ with $c_{f}$.  
%The difference between $c_{0}$ and $c_{i}$ is the water lost during heating, and that between $c_{i}$ and $c_{f}$ the water lost during the measurement itself; both are obtained by weighing.}
%\label{tab:exp_conditions}
%\setlength{\tabcolsep}{2pt}
%\small
%\begin{tabular}{cccccc}
%\hline
%$c_{0}$ (mM) & $c_{0}/\cmc_{25}$ & $c_{i}$ (mM) &
%$T_\mathrm{max}$ ($\degC$) & $c_{f}$ (mM) & $T_\mathrm{min}$ ($\degC$) \\
%\hline
% 0.16 & 0.02 &  0.17 & 73.5 &  0.18 & 25.4 \\
% 0.41 & 0.05 &  0.43 & 80.1 &  0.46 & 25.5 \\
% 0.57 & 0.07 &  0.60 & 83.5 &  0.64 & 25.6 \\
% 0.82 & 0.1 &  0.87 & 76.4 &  0.93 & 24.8 \\
% 1.64 & 0.2 &  1.71 & 81.0 &  1.81 & 25.6 \\
% 2.46 & 0.3 &  3.04 & 81.7 &  3.26 & 25.9 \\
% 4.10 & 0.5 &  4.38 & 82.1 &  4.65 & 25.3 \\
% 8.20 & 1 &  8.55 & 76.7 &  8.95 & 25.6 \\
% 9.02 & 1.1 &  9.28 & 76.0 &   --- & 27.5 \\
% 9.84 & 1.2 & 10.2 & 76.1 & 10.91 & 25.7 \\
%10.7 & 1.3 & 11.1 & 82.4 & 11.57 & 25.7 \\
%12.3 & 1.5 & 13.3 & 83.7 & 13.92 & 25.5 \\
%16.4 & 2 & 16.9 & 75.7 & 17.65 & 25.7 \\
%24.6 & 3 & 25.2 & 77.6 & 26.40 & 25.7 \\
%32.8 & 4 & 33.4 & 77.0 & 35.27 & 25.5 \\
%41.0 & 5 & 42.9 & 78.5 & 45.54 & 25.6 \\
%\hline
%\end{tabular}
%\end{table}

\begin{table}[t]
\centering
\caption{Solutions examined in this work.  
Each solution was made up at the nominal concentration $c_{0}$, heated, and then measured while cooling, so that the run begins at $T_\mathrm{max}$ with concentration $c_{i}$ and ends at $T_\mathrm{min}$ with $c_{f}$.  
The difference between $c_{0}$ and $c_{i}$ is due to the water loss during heating, and that between $c_{i}$ and $c_{f}$ the water loss during the measurement itself; both are obtained by weighing.}
\label{tab:exp_conditions}
\setlength{\tabcolsep}{8pt}
\begin{tabular}{cccc}
\hline
$c_{0}$ & $c_{0}$ &
$c_{i} \rightarrow c_{f}$ &
$T_\mathrm{max} \rightarrow T_\mathrm{min}$ \\
($\cmc_{25}$) & (mM) & (mM) & ($\degC$) \\
\hline
0.02 &  0.16 &  0.17 $\rightarrow$  0.18 & 73.5 $\rightarrow$ 25.4 \\
0.05 &  0.41 &  0.43 $\rightarrow$  0.46 & 80.1 $\rightarrow$ 25.5 \\
0.07 &  0.57 &  0.60 $\rightarrow$  0.64 & 83.5 $\rightarrow$ 25.6 \\
0.1  &  0.82 &  0.87 $\rightarrow$  0.93 & 76.4 $\rightarrow$ 24.8 \\
0.2  &  1.64 &  1.71 $\rightarrow$  1.81 & 81.0 $\rightarrow$ 25.6 \\
0.3  &  2.46 &  2.58$^{*}$ $\rightarrow$  2.76 & 81.7 $\rightarrow$ 25.9 \\
0.5  &  4.10 &  4.38 $\rightarrow$  4.65 & 82.1 $\rightarrow$ 25.3 \\
1.   &  8.20 &  8.55 $\rightarrow$  8.95 & 76.7 $\rightarrow$ 25.6 \\
1.1  &  9.02 &  9.28 $\rightarrow$  9.7$^{+}$  & 76.0 $\rightarrow$ 27.5 \\
1.2  &  9.84 & 10.2 $\rightarrow$ 10.9 & 76.1 $\rightarrow$ 25.7 \\
1.3  & 10.7  & 11.1 $\rightarrow$ 11.6 & 82.4 $\rightarrow$ 25.7 \\
1.5  & 12.3  & 13.3 $\rightarrow$ 13.9 & 83.7 $\rightarrow$ 25.5 \\
2.   & 16.4  & 16.9 $\rightarrow$ 17.7 & 75.7 $\rightarrow$ 25.7 \\
3.   & 24.6  & 25.2 $\rightarrow$ 26.4 & 77.6 $\rightarrow$ 25.7 \\
4.   & 32.8  & 33.4 $\rightarrow$ 35.3 & 77.0 $\rightarrow$ 25.5 \\
5.   & 41.0  & 42.9 $\rightarrow$ 45.5 & 78.5 $\rightarrow$ 25.6 \\
\hline
\multicolumn{4}{l}{\footnotesize $^{*}$Estimated; the mass before the run was not recorded.}\\
\multicolumn{4}{l}{\footnotesize $^{+}$Estimated; the mass after the run was not recorded.}\\
\end{tabular}
\end{table}
%%%%%%%%%%
\subsection{Protocol}

Once the solution is prepared, the measurement proceeds as follows.
First, we fill the glass container with 100\,mL of prepared solution and heat it up using a hot plate.
During the heating process, which takes approximately 10 to 20 minutes, evaporation takes place, changing the concentration from $c_0$ to $c_i$.
We measure the solution weight before and after the heating, and on average $4.3\pm1.7$\% of water is lost during the heating.
Then we quickly move the container to the tensiometer.
Second, we set up the thermocouple to measure the temperature of solution and start the measurements while the solution is set to cool down slowly in the chamber.
Depending on how fast we move the container from the balance to the tensiometer, the starting temperature $T_\mathrm{max}$ varies from run to run.
In each measurement, the surface tension $\sigma$ is taken as the maximum of the force reading, and the temperature $T$ is recorded at that instant.
The surface tension measurement is repeated as frequently as possible at high temperature where the temperature changes most drastically, but at low temperature we measure in 2\,$\degC$ intervals.
Finally, when the solution is cooled down to room temperature, we measure its weight again. 
The evaporation during the measurement is approximately $5.2\pm0.8$\%.
The final concentration $c_f$ is therefore on average 5.5\% higher than $c_i$: in each run the low-temperature data are taken at a higher concentration than the high-temperature data.
In Table~\ref{tab:exp_conditions}, we summarize the experimental conditions.

We note that all components in contact with the solution, including the glass container and the Pt plate, were thoroughly cleaned before each run: cleaned in an ultrasonic bath for 3 minute, and rinsed with DI water. 
Measurements were always carried out from low to high concentration, so that any carry-over from dilute solutions, if any survived the cleaning process, would not yield a substantial effect on more concentrated solutions.
The Pt plate was flamed once at the beginning of the experiments.

%%%%%%%%%%
\subsection{Validation, uncertainty and scope of the data}

Our preliminary characterization using DI water shows that the total measurement error is approximately 2\%.
We believe the error mostly comes from the temperature measurement for two reasons.
First, the temperature drops quickly at the high end of the range, at about 0.1\,$\degC/\mathrm{s}$.  
A single measurement takes roughly 30\,s, so the temperature can drift by 2 to 4\,$\degC$ while it is being made.
Even though we standardize the temperature to be measured at the moment of the maximum force reading, the high variability of temperature leaves a moderate uncertainty.
%For consistency, we deliberately record the temperature at the moment of detachment, but the high variability in temperature leaves us a moderate uncertainty.
Second, we are measuring the temperature of bulk fluid in the container rather than the surface temperature.
Because the solution cools down by interacting with air, the surface is expected to be cooler than the bulk, causing a systematic error.
In this work, we did not quantify these two errors individually, but the DI water result falls within a $\pm$2\% band of the established measurement from International Association for the Properties of Water and Steam (IAPWS) \cite{iapws2014}.
%indicates that they together yield approximately 2\% uncertainty.
%Considering that without heating the tensiometer consistently returns 72\,mN/m at room temperature, we presume that there is no noticeable error in our tensiometer.
At room temperature, the instrument consistently yielded approximately 72 mN/m for DI water, suggesting that instrumental bias is small compared with the temperature-related uncertainties discussed above.

%%%%%%%%%%%%%%%%%%%%%%%%%%%
\section{Result}
%%%%%%%%%%%%%%%%%%%%%%%%%%%

%%%%%%%%%%
\subsection{Surface tension versus temperature}

\begin{figure}
\centering
\includegraphics[width=0.6\columnwidth]{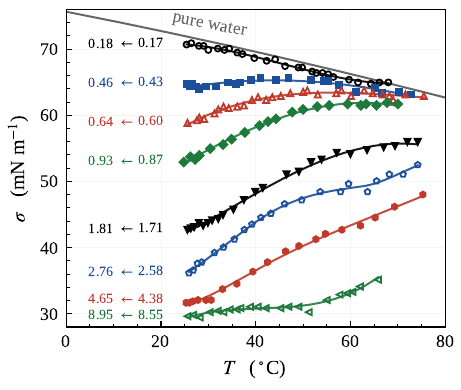}
\caption{Surface tension of aqueous SDS solutions with respect to the temperature, for eight concentrations at and below the \cmc.
Labels give the concentration in mM at the end and at the start of each run ($c_f\leftarrow c_i$); the difference is the evaporative enrichment during the measurement.  
The grey line is the IAPWS data for pure water~\cite{iapws2014}.  
Above about 0.4\,mM the slope is positive, opposite in sign to that of water.  
Curves through the points are smoothing splines and are drawn for guidance only.
}
\label{fig:sigma-vs-T}
\end{figure}

In Fig. \ref{fig:sigma-vs-T}, our measurements of surface tension $\sigma$ are shown with respect to the temperature $T$ for several concentrations, ranging $c_i=0.17$\,mM to 8.55\,mM.
At $c_i=0.17\,{\rm mM}$ (0.02 $\cmc_{25}$), $\sigma$ decreases as $T$ increases in the entire test range, just like pure water~\cite{iapws2014}.
Already at $c_i=0.43\,\rm mM$, however, this trend is reversed. 
At this concentration, $\sigma$ is nearly independent of $T$, however as the concentration increases the low temperature surface tension decreases while the high temperature surface tension remains relatively unchanged.
At $c_i>0.87\,\rm mM$, it is evident that $\sigma$ actually increases as $T$ increases.
The average slope is maximized near $c_i=2.58\,\rm mM$, but as the concentration approaches \cmc, the slope falls off sharply.

%%%%% FIGURE 1 %%%%%

In the figure, it is seen that the effect of surfactant that decreases the surface tension weakens as $T$ increases.
Near room temperature, $\sigma$ differs by about 40\,mN/m between pure water and the solution at \cmc, but this difference diminishes as $T$ increases.
It even appears that $\sigma\rightarrow\sigma_0$ asymptotically, where $\sigma_0$ is the surface tension of pure water.
This indicates the reduction of surfactant adsorption at the surface.

%%%%%%%%%%
%\subsection{Surface tension above the \cmc}

\begin{figure}
\centering
\includegraphics[width=0.6\columnwidth]{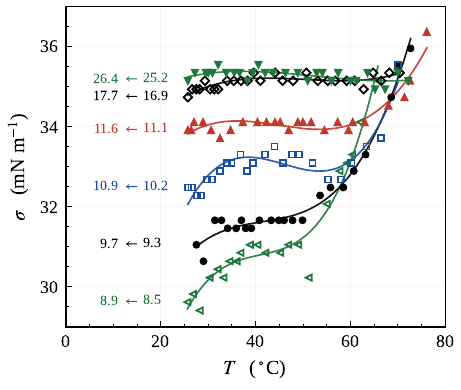}
\caption{Surface tension with respect to the temperature, for six concentrations at and above the \cmc.
Labels denote the end and the start concentrations ($c_f\leftarrow c_i$). 
Three high-concentration solutions (13.3, 33.4 and 42.9\,mM) are not shown in this figure because they overlap with 25.2\,mM data.
Also, data at $c_i=8.5\,\rm mM$ are the same as those in Fig. \ref{fig:sigma-vs-T} but with expanded ordinate.
}
\label{fig:sigma-vs-T-highconc}
\end{figure}

Figure \ref{fig:sigma-vs-T-highconc} shows the measurement of $\sigma(T)$ at higher concentrations, i.e. $c_i\geq 8.55 \,\rm mM$. 
In the figure, it is observed that the temperature dependence weakens as the concentration surpasses \cmc.
Slightly above \cmc, for example at $c_i=9.3\,\rm mM$, $\sigma$ ranges from $30.6\,\rm mN/m$ to $32.5\,\rm mN/m$ over the wide range.
However when $c_i>16.9\,\rm mM$, $\sigma$ is almost constant at $35.2\pm0.3\,\rm mN/m$.
Although not shown in the figure, data at 13.3, 33.4, 42.9\,mM overlap with 16.9 and 25.2\,mM data, supporting further that the surface tension is insensitive to the temperature in the high concentration regime.

%%%%% FIGURE 2 %%%%%

\begin{figure}
\centering
\includegraphics[width=0.6\columnwidth]{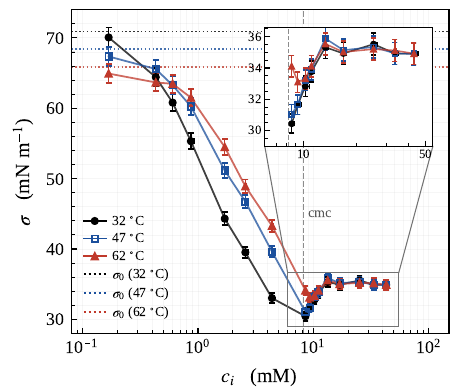}
\caption{
Surface tension of aqueous SDS solutions with respect to concentration at three temperatures, interpolated from all sixteen runs of Table~\ref{tab:exp_conditions}.
%This figure is created by interpolating data in Figs. \ref{fig:sigma-vs-T} and \ref{fig:sigma-vs-T-highconc}. 
Vertical error bars show 2\% uncertainty in measuring temperature, and the horizontal error bars showing up to 5\% uncertainty in concentration are smaller than the point itself.
Dotted lines show the surface tension of pure water~\cite{iapws2014} at each temperature, and the dashed line marks $\cmc_{25}$.
Inset: the region around the \cmc.}
\label{fig:isotherm}
\end{figure}

In this regime, it is notable that $\sigma$ actually increases as the concentration increases.
This may seem to be counter-intuitive, but it is a well-known effect of the interaction between micelle and dodecanol impurity~\cite{miles1944,vollhardt2000,kralchevsky2003,correia2022}.
Dodecanol is a byproduct that is present as an impurity in SDS and reduces the surface tension of SDS solutions.
Over \cmc, SDS molecules form micelles, which remove dodecanol from the surface.
This effect is better visualized by plotting $\sigma$ with respect to  $c_i$.
In Fig. \ref{fig:isotherm}, $\sigma$ decreases with $c_i$, but it bounces back near \cmc\ to approach a saturation value $\sigma_{sat}=35.2\,\rm mN/m$, creating a `dip' of approximately 5\,mN/m.
Because $\sigma_{sat}$ is considered as the surface tension without impurity, from the depth of this `dip', the impurity level can be gauged.
Comparing with data in the literature~\cite{blokhuis2025}, we estimate that our dodecanol impurity is approximately $ 0.2\%$.

%%%%% FIGURE 3 %%%%%

To wrap up, our data show that SDS solutions exhibit a strong temperature dependence of surface tension below \cmc\ but virtually none above \cmc.
Potentially, this is a useful control parameter for future fluid experiments. 
One should use concentrations below \cmc\ if interested in thermocapillary flows. 
However, if one is interested in using soap films for two-dimensional Rayleigh-B\'enard convection, concentrations above \cmc\ may be exploited.

%%%%%%%%%%
\subsection{The temperature dependence of c.m.c.}

In this section, we check if the observed temperature dependence is due to the temperature dependence of \cmc.
Below the cmc there are no micelles, so the \cmc\ does not enter the adsorption equilibrium at all. 
We nevertheless test this scenario, because the temperature dependences of $\sigma$ and \cmc\ both run in the same direction and invite the explanation.

It is widely acknowledged that \cmc\ depends on the temperature, and it has been measured by conductometry~\cite{goddard1957,flockhart1961,chatterjee2001} and by titration calorimetry~\cite{chatterjee2001,paula1995}.
They commonly report that \cmc\ is minimized between 20 and 35\,$\degC$ at $\sim$8.2\,mM and increases on either side.
They report \cmc\ to be 10\,mM at 60\,$\degC$ and up to 13\,mM at 80\,$\degC$.% \cite{paula1995}.

First, we estimate \cmc\ by using our surface tension data. 
We use the fact that the location of the `dip' due to the dodecanol impurity marks \cmc.
By assuming that near the dip $\sigma$ is a quadratic function of $\ln c_i$, we obtain a rough estimate of the \cmc\ at those temperatures where the minimum lies within our concentration range.

Second, we assume that the temperature dependence of the surface tension discussed above is entirely due to $\cmc(T)$, and estimate the \cmc\ on that basis.  
This amounts to finding the factor by which the concentration axis must be rescaled to bring the curves of $\sigma_0-\sigma$ onto a single one.
%Nevertheless this attempt is not really successful because we can `force' the collapse only for the small concentration regime, but we can get the estimate based on the partial collapse.
In fact the attempt is not entirely unsuccessful.
The collapse can be forced only below about 4\,mM, but that partial collapse is enough to give an estimate.

\begin{figure}
\centering
\includegraphics[width=0.6\columnwidth]{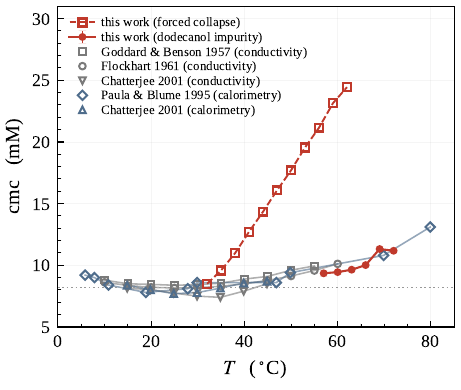}
\caption{
Critical micelle concentration of SDS in water versus temperature.
Filled red circles are the position of the surface tension minimum in Fig.~\ref{fig:isotherm}, obtained by using the surface tension dip due to the dodecanol impurity, and this result is consistent with previous measurements based on electrochemical and calorimetric methods.
Open red squares are obtained by assuming that the temperature dependence of $\sigma$ is entirely due to $\cmc(T)$, and this result deviates from the other measurements.
Grey and blue symbols are literature values, measured by conductivity~\cite{goddard1957,flockhart1961,chatterjee2001} and by titration calorimetry~\cite{chatterjee2001,paula1995}.
The dotted line marks $\cmc_{25}=8.2$\,mM.
}
\label{fig:temperature-dependence-cmc}
\end{figure}

Figure \ref{fig:temperature-dependence-cmc} summarizes previous measurements~\cite{goddard1957,flockhart1961,chatterjee2001,paula1995}, our estimate using the dodecanol impurity, and our other estimate based on the assumption that the temperature dependence of $\sigma$ is entirely due to $\cmc(T)$. 
It is obvious that the estimation based on the assumption deviates from other measurements and is even not physical, indicating that the assumption is incorrect.
%It is seen that our measurement using the impurity matches well with previous electrochemical measurements, suggesting that our solutions are well-prepared.
On the other hand, the estimate from the impurity dip follows the literature values, so the failure of the second estimate cannot be attributed to our solutions having an unusual \cmc.

%%%%% FIGURE 4 %%%%%

%%%%%%%%%%
\subsection{Temperature coefficients}

For quantitative analyses, we define the temperature coefficient $\sigma'$ as the first derivative of $\sigma$ with respect to $T$.
We include the correction of evaporation effect during experimental runs as follows.
\begin{equation}
\sigma'(T)=\left(\frac{\partial\sigma}{\partial{T}}\right)_c=\frac{d\sigma}{dT}-\left(\frac{\partial\sigma}{\partial{c}}\right)_T\cdot \frac{dc}{dT}.
\label{eq:def_temp_coeff}
\end{equation}
The fractional evaporation rate $d\ln{c}/dT=-1.17\times 10^{-3}\,\rm{K^{-1}}$ is common for all experimental conditions, and $(\partial\sigma/\partial{c})_T$ is evaluated from experimental data.
The correction term, the second term in Eq. \eqref{eq:def_temp_coeff} is approximately 3\% of $d\sigma/dT$.
Therefore, our temperature coefficient represents the response of $\sigma$ on the variation of $T$ at fixed concentration.

%For quantitative analyses, we define and present the temperature coefficient $\sigma'$ of surface tension.
%When $\sigma$ is expanded with respect to $T$ at fixed concentration,
%\begin{equation}
%\sigma(T)=\sigma(T_0)+\delta T \left(\frac{\partial\sigma}{\partial T}\right)+ \mathcal{O}(\delta T^2),
%\end{equation}
%where $\delta T = T-T_0$.
%We take the {\it local} temperature coefficient $\sigma'\equiv d\sigma/dT$, evaluated at $T_0$.

\begin{figure}
\centering
\includegraphics[width=0.6\columnwidth]{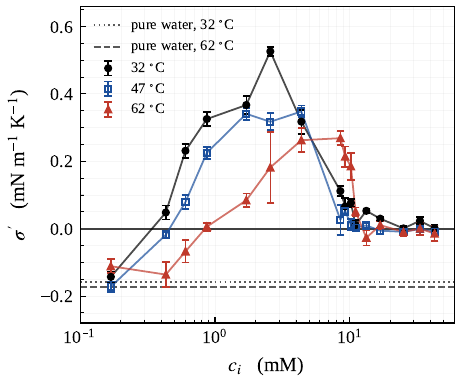}
\caption{Temperature coefficient $\sigma'$ versus concentration at three temperatures.
At the low concentration, it has a value similar to that of water, $-0.17\,\rm mN/m/K$ (at 47\,$\degC$).
However, it has a positive value in the range of $0.3\lesssim c_i<8$\,mM, meaning that the surface tension increases with the temperature.
The absolute value of temperature coefficient can be thrice as large as that of pure water.
Dotted and dashed lines show $\sigma'$ of pure water~\cite{iapws2014} at 32\,$\degC$ and 62\,$\degC$, respectively.
}
\label{fig:dsigmadt}
\end{figure}

Fig. \ref{fig:dsigmadt} shows the calculation of $\sigma'$ with respect to $c_i$ for three different temperatures, 32, 47, and 62\,$\degC$.
These data are calculated by taking the slope of measured points within a $\pm 8$\,K window of the stated temperature. 
At all temperatures, when $c_i$ is small, $\sigma'$ approaches that of water, and this is also consistent with the measurement in Ref. \cite{granada1998,judd2005} where they reported $\sigma'$ is indistinguishable from that of water below 0.14 mM of SDS.
As $c_i$ increases, $\sigma'$ changes the sign from negative to positive.
The reversal of sign in $\sigma'$ was reported in SDS solution with hexane~\cite{motomura1978} and in other alkyl dodecyl sulfates~\cite{lu1993}.
We estimate the concentration $c_x$ at the cross-over, by taking the $x$-intercept in Fig. \ref{fig:dsigmadt}, and it depends on $T$, in other words, $c_x=c_x(T)$.
After the cross-over, $\sigma'$ reaches its maximum value $\sigma'_{\rm max}$ and decreases back to zero near \cmc.

%%%%% FIGURE 5 %%%%%

\begin{figure}
\centering
\includegraphics[width=0.6\columnwidth]{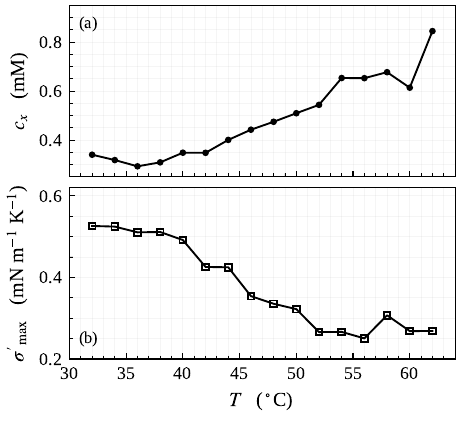}
\caption{
Two features of $\sigma'(c)$.
(a) The concentration $c_x$ at which $\sigma'$ changes sign.
(b) The largest value of $\sigma'$ at each temperature.}
\label{fig:c_x}
\end{figure}

In Fig. \ref{fig:c_x}, $c_x$ and $\sigma'_{\rm max}$ are presented.
We find that $c_x$ increases with $T$ and that $\sigma'_{\rm max}$ decreases with $T$.
Both observations suggest that the effect of surfactant is weakened at high temperature.
Other notable features include the followings: 
First, it only takes 0.3 to 0.8 mM to flip the sign of $\sigma'$.
These concentrations are approximately 0.05 to 0.1 \cmc\ at room temperature.
Second, $\sigma'_{\rm max}$, for $T<38\,\degC$, is greater than 0.5\,mN/m/K. 
This number is three-fold greater than that of pure water in its amplitude.
The thermocapillary effect in SDS solutions not only flips its sign but is also very strong.

%%%%% FIGURE 6 %%%%%

The temperature coefficient is a thermodynamic quantity.
From Eq. \eqref{eq:Gibbs-Duhem}, we get
\begin{equation}
%s=-\frac{d\sigma}{dT} - \Gamma \frac{d\mu}{dT},
\sigma'=\frac{d\sigma}{dT} =-s- \Gamma \frac{d\mu}{dT},
\label{eq:entropy}
\end{equation}
where $s$ is the entropy per unit area, $\Gamma=N/A$ is the surface concentration of surfactant, $\mu$ is the chemical potential.
%In general, $\Gamma$ is related with the bulk concentration $c$ through the Langmuir adsorption isotherm, 
%\begin{equation}
%\Gamma=\Gamma_\infty \frac{1}{1+a/c},
%\label{eq:Langmuir}
%\end{equation}
%where $\Gamma_\infty$ is the saturation value and $a$ is the half-saturation concentration.
The temperature coefficient can be positive because $d\mu/dT$ is negative. 
The chemical potential, which is the molar Gibbs free energy, is largely accounted for by the entropy of dispersal which is large and positive for a dilute solute. 
Therefore, if $-\Gamma d\mu/dT$ is greater than the surface entropy $s$, the temperature coefficient becomes positive.

Finally, we can estimate $\Gamma$ with our data.
By using 
\begin{equation}
\Gamma=-\frac{1}{\nu RT} \frac{d\sigma}{d\ln c},
\label{eq:Gamma}
\end{equation}
where $R$ is the ideal gas constant and $\nu=2$ is the Gibbs prefactor.
We find that $\Gamma$ averaged over the range $2.58\leq c_i \leq4.38\,\rm mM$ decreases from 2.4 to 1.9\,$\rm \mu mol/m^2$ as $T$ increases from 32 to 62\,$\degC$, as shown in Fig. \ref{fig:speculation}.
Considering that $\Gamma$ at 4.38\,mM is 80\% to 90\% of $\Gamma_{\cmc}$ \cite{xu2013}, we get $\Gamma_\cmc$ is approximately 3\,$\rm \mu mol/m^2$.
This number is about 20\% lower than the values reported in the literature \cite{lu1993,wilson1957,elworthy1966}.

%%%%%%%%%%
\subsection{Speculation on Langmuir adsorption isotherm}

The surface concentration $\Gamma$ can also be estimated from the Langmuir adsorption isotherm \cite{langmuir1918},
\begin{equation}
\Gamma=\Gamma_\infty \frac{1}{1+a/c},
\label{eq:Langmuir_adsorption_isotherm}
\end{equation}
where $\Gamma_\infty$ and $a$ are the parameters of the model.
To determine these parameters as functions of temperature, we assume $a=\cmc/\alpha$, where $\alpha$ is a constant independent of concentration and temperature.
We approximate $\cmc(T)=8.2+k(T-25)^2$ with $k=0.00122\,\rm mM/K^2$, the best fit to our own data in Fig.~\ref{fig:temperature-dependence-cmc}.
The nonlinear regression returns $\alpha=4.88\pm0.86$; from 32 to 62\,$\degC$, $a$ varies from 1.69 to 2.02\,mM and $\Gamma_\infty$ from 6.38 to 3.40\,$\rm \mu mol/m^2$.
We note that 6.38\,$\rm \mu mol/m^2$ is too large compared with the reported values of around 4\,$\rm \mu mol/m^2$ \cite{lu1993,xu2013}, as it would leave only 0.26\,$\rm nm^2$ per surfactant molecule.
A possibility remains that $\Gamma_\infty$ is actually greater than the normal range due to dodecanol \cite{xu2013}.

\begin{figure}
\centering
\includegraphics[width=0.6\columnwidth]{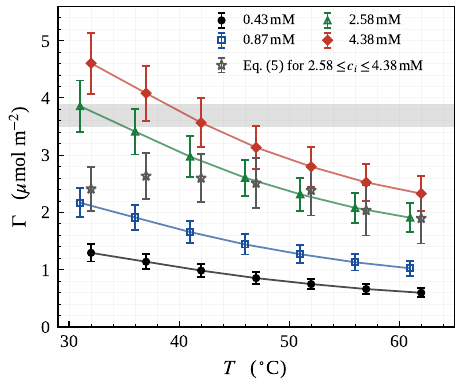}
\caption{
The calculated $\Gamma$ by using Eq. \eqref{eq:Gamma} (black stars) and Eq. \eqref{eq:Langmuir_adsorption_isotherm} (other symbols).
The gray band remarks the reported value of $\Gamma_\infty$~\cite{lu1993,xu2013}.
Data for 0.87\,mM and 2.58\,mM are shifted by 1\,$\degC$ for visibility.
}
\label{fig:speculation}
\end{figure}

We then calculate $\Gamma$ at several $c_i$ using the fitted $\Gamma_\infty(T)$ and $a(T)$, as shown in Fig.~\ref{fig:speculation}.
Although $\Gamma$ is evidently overestimated, the decreasing trend with temperature is clear.
This outcome is speculative and rests on an assumption, but it supports the view that the surfactant tends to leave the surface as the temperature increases.

%%%%% FIGURE 7 %%%%%

%%%%%%%%%%%%%%%%%%%%%%%%%%%
\section{Conclusions}
%%%%%%%%%%%%%%%%%%%%%%%%%%%

To conclude, we measured the surface tension of aqueous sodium dodecyl sulfate solutions at concentrations from 0.17 to 43\,mM and temperatures from 26 to 83\,$\degC$.
We find that the temperature coefficient is positive, i.e., $d\sigma/dT>0$, for the concentration range from $c_x$ to \cmc, where $c_x$ has been measured to be 0.3 to 0.8\,mM, approximately a tenth of \cmc\ at room temperature.
At its maximum, the temperature coefficient reaches $+0.53\,\rm mN/m/K$, thrice that of water in magnitude.

Effectively, the raising the temperature dilutes the surface.
Our result cannot be attributed to the temperature dependence of \cmc, as the effect is far stronger than what is expected from it.
We believe that at high temperature the bulk is favored over the surface because of the entropy of dispersal. 

Our results can be used for future fluid experiments that involve the thermocapillary effect.
We do not claim the measurements presented here are the properties of pure sodium dodecyl sulfate; we deliberately used the sample of 97\% purity as received, reflecting the reality in fluid laboratories across the world.
Based on the results presented here, the direction of thermal response can be adjusted by changing the concentration and temperature.

\bibliographystyle{apsrev4-2}
\bibliography{surface-tension-temperature-2026}
\end{document}